\documentclass[epj]{svjour}
\usepackage{graphics}
\usepackage{graphicx}
\usepackage{epsfig}
\begin{document}
\title{Revisiting the nonequilibrium phase transition of the triplet-creation model}
\author{Giovano O. Cardozo and 
	Jos\'e F. Fontanari \thanks{e-mail: {\tt fontanari@ifsc.usp.br} }
       }                    
\institute{Instituto de F\'{\i}sica de S\~ao Carlos,
Universidade de S\~ao Paulo, Caixa Postal 369, 13560-970 S\~ao Carlos, 
S\~ao Paulo, Brazil}
\date{Received: date / Revised version: date}
%
\abstract{
The nonequilibrium phase transition in the triplet-creation model is investigated
using critical spreading and the conservative diffusive contact process. The results
support the claim that at high enough diffusion the phase transition becomes discontinuous.
As the diffusion probability increases the critical exponents change continuously from the
ordinary directed percolation (DP) class to the compact directed percolation (CDP). The fractal
dimension of the critical cluster, however, switches abruptly between those two universality
classes.   Strong crossover effects in both methods make it difficult, if not impossible, to
establish the exact location of the tricritical point.
} 
\titlerunning{The triplet-creation model revisited}
\maketitle
\section{Introduction} \label{sec:Intro}

The occupation of a resident population by a small colony of mutant replicators or, more simply, the settlement of a colony
of spontaneously generated replicators in the vacuum are critical issues in prebiotic evolution \cite{Maynard}, 
which have recently been shown to be nonequilibrium phase transitions \cite{Ferreira}. The 
characterization of the diverse types of replicators according to the conditions necessary to their replications 
\cite{Michod} leads to a variety of irreversible  dynamical systems that are familiar to  the statistical physics
community \cite{Marro,review,Odor}. Of particular importance for the setting of a sound prebiotic scenario is the 
nature of the invasion process, that can be suitably described by the dynamic and static critical exponents 
associated to the probability of invasion \cite{spreading}.

In this contribution we re-examine the one-dimensional  triplet-creation model of Dickman and Tom\'e \cite{DT},
in which  a necessary condition for replication (i.e., particle creation)  is the existence of 
at least three replicators occupying contiguous positions in the neighborhood of an empty site. 
(Henceforth we will use the terms replicator and particle interchangeably.)
This may be viewed as a generalization of the three-member hypercycle \cite{Hyper,Boerlijst} 
in which the presence of all members is required to catalyze the replication of any of 
the component replicators.
Decay and diffusion of the replicator to  neighboring sites are considered as well. 
The rich and controversial  critical behavior  of the triplet-creation model owns to the 
competition between diffusion and the triplet replication process.  The motivation for the proposal 
of that model was to find the simplest, \textit{local}
nonequilibrium model that exhibits a discontinuous transition into an absorbing state \cite{DT}.

In the absence of diffusion,  the triplet-creation model exhibits a
continuous transition to the unique absorbing vacuum state which
belongs to  the class  of universality of the directed percolation, DP \cite{Marro,review,Odor}, as predicted
by  a conjecture put forward independently by Janssen and Grassberger \cite{DP,Grassberger}.
When diffusion is turned on, however, the situation becomes much less clear.
The original analysis of Dickman and Tom\'e \cite{DT} indicates that the continuous transition 
associated to the low diffusion regime changes into a discontinuous transition for high enough diffusion. (Of course,
for infinite diffusion rate, and infinite number of particles as well, the  mean-field limit is reached and so
the transition is definitely discontinuous.) 
Their conclusion  was disputed by Hinrichsen  \cite{Haye} who presented general arguments against the existence
of discontinuous transitions in certain classes of one-dimensional irreversible models and provided numerical
evidence, using spreading analysis, that the transition of the triplet-creation model is always continuous, 
regardless
of the value of the diffusion rates. Moreover, the transition was found to be in the DP universality class.
More recently, Fiore and de Oliveira \cite{Fiore} used 
the conservative diffusive contact process, in which the number of particles is kept fixed (see also \cite{CCP}),
to vindicate the original findings of Dickman and Tom\'e.
Here we investigate the nonequilibrium transition of the triplet-creation model using both the spreading and
the conservative diffusive techniques and find, in agreement with \cite{Fiore} strong evidence of 
a discontinuous transitions in the high diffusion regime.  However, strong crossover effects lead to
a continuous variation of the dynamic and static critical exponents, which obey the generalized hyperscaling
relation \cite{Mendes},  so that a precise location
of the tricritical point (i.e., the point at which the transition becomes discontinuous) is
very difficult. These crossover effects are probably the cause of the controversies surrounding the triplet-creation
model.

The rest of the paper is organized as follows. In Sect.~\ref{sec:Model}  we present the 
set of rules that govern the evolution of the particles in the triplet-creation model
with a slight variation from the original proposal.
The two techniques  used to characterize the stationary and  dynamic behavior at
the transition point are reviewed in Sect.~\ref{sec:Methods}. In particular, we give emphasis to
the description of the  conservative diffusive contact process since it is, relatively to the
spreading analysis, a new method to study   nonequilibrium phase transitions. The results obtained by the
application of  both methods to the triplet-creation model are then presented and discussed
in Sect.~\ref{sec:Results}. Finally, in  Sect.~\ref{sec:Conclusion} we  present 
some concluding remarks.

\section{The triplet-creation model} \label{sec:Model}
To gain in computation speed, instead of describing the configuration of the $L$-sites chain  
in terms of binary occupation variables \cite{DT}, we choose to describe the chain
by a list of the particle positions $p_n = 1, \ldots, L$ with $n=1, \ldots, N$. (In practice, 
a second list $q_i = 0,1, \ldots, N$ with $i=1, \ldots, L$ is needed in order to identify which particle occupies
site $i$. If $q_i = 0$ then site $i$ is vacant.) This procedure was also used in \cite{Haye}
for without it long time spreading simulations would not have been possible.
In terms of these lists the evolution rules are as follows
(see \cite{Dick} for a similar approach). 

\textit{Diffusion}: This is the diffusion or hopping process that occurs with probability $D$.
A particle, say $n$, is chosen at random and one of its nearest neighbors sites, say $k$, is chosen also 
at random. If $q_k=0$ then particle $n$ moves to site $k$ and the lists are updated accordingly, otherwise
nothing happens. 

\textit{Decay}: This process occurs with probability 
\begin{equation}\label{gamma}
\gamma = \frac{ 1 - D}{1 + \lambda} .
\end{equation}
A randomly chosen particle, say $n$, is chosen and then annihilated. This is implemented by simply moving
particle $N$ to the position of particle $n$, with the corresponding changes in the lists, and then
by resetting the number of particles to $N-1$.

\textit{Creation}: This process takes place with probability $s = \lambda \gamma$. As before, a particle,
say $n$ with $p_n = i$, is chosen at random. Then its two nearest-neighbor sites, $i-1$ and $i+1$, are checked
to verify whether they are both occupied  or not. If $q_{i-1} q_{i+1} = 0$ nothing happens, since then particle
$n$ will  not be part of a triplet,  otherwise one of
the sites $i-2$ or $i+2$ is chosen at random and a new particle, identified by the label $N+1$, 
is placed at the chosen site, provided
it is vacant. 
The number of particles is then reset to  $N+1$, so the newly created particle is placed
at  the end of the $q_i$ list.

Since $D+\gamma + s = 1$ there are only two independent parameters in the model which we choose
as $D$ and $\gamma$. As we will see in Sect.~\ref{sec:Methods}, the fact that the probability of decay and creation 
are not independent introduces some
complications to the formulation of the conservative diffusive contact process for the
triplet-creation model. 

The modifications  we have introduced in the processes of decay and creation were so as to reduce 
the number of wasted checks of the original formulation of the model, such as choosing an empty site in any
of these two pro\-cesses. Hence they have
no effect whatsoever on the steady-state properties of model. As for the dynamics, we acknowledge that by increasing
the number of trials that effectively modify the chain configuration, we may
affect in a nontrivial and uncontrollable way the time dependence of some properties of the
model. But we do not expect that these effects will alter the asymptotic
form of the relevant dynamic quantities,  say from exponential to power-law and vice-versa, or
the values of the exponents that characterize an eventual power-law decay; otherwise the current effort to
classify nonequilibrium phase transitions \cite{Marro,review,Odor} would be utterly vain.

The changes in the hopping process, however, are a different matter. In the original model, a site 
chosen at random, say $i$, is interchanged with its right neighbor, site $i+1$ \cite{DT}. (If both
sites are occupied or vacant, then their interchange  will be a waste of time, which is avoided in our
formulation.)  Hence in the original model a particle at site $i$ has two opportunities to diffuse -- 
when either site $i$ or 
site $i+1$ is chosen -- but it has only one chance of hopping in the present framework. There is, fortunately,  
a simple relationship between the probability of hopping $\tilde{D}$ of the original
model and our parameter $D$, namely, $\tilde{D} = D/ \left ( 2 - D \right )$  \cite{Dick}, so
our results  can be readily compared with those in the literature.

For the sake of concreteness we define a  trial as the choice of one of the three processes -- diffusion, decay and
creation -- described above. In the original algorithm,  a time step of the dynamics is defined as the realization of $L$ 
such a trials \cite{DT}, which is clearly impracticable in the case  of the very large (ideally infinite)   chains
used in the spreading analysis. Instead, we follow Ref.\ \cite{Dick} and define the time increment per trial 
as $1/N$, where $N$ is the number of particles just before the trial.  Hence each trial represents on average
$L/N$  trials in the original algorithm. 

\section{Methods} \label{sec:Methods} 

In what follows we present a brief account of the spreading analysis which is based on the time
evolution of the model as defined in the previous section. In particular, the number of particles varies in
time following the separated creation and annihilation processes. More emphasis is given to the description
of the, comparatively less familiar,  conservative diffusive contact process, in which the number of particles
is kept fixed during the evolution of the colony.

\subsection{Spreading analysis} 

We begin with the spreading analysis \cite{spreading} since it is probably the simplest and most powerful 
technique  to estimate
the values of the critical parameters at which the transition between the active and
the absorbing regimes take place. We set an initial colony of replicators -- a single triplet for $D=0$
and a string of $40$ contiguous replicators for $D> 0$  --
in the center of an otherwise empty cell of ``infinite'' size. This can be
accomplished by taking the chain size large enough so that, during the time we follow
the evolution of the colony, the replicators can never reach the chain extremes.
Here we focus on the time dependence of three key quantities: (i) the average  number of
replicators $N(t)$; (ii) the survival probability of the colony $P(t)$; and 
(iii) the average mean-square distance over which the colony has spread $R^2(t)$. For each
time $t$ we carry out  $10^5$  independent runs, all starting with the same colony. Hence
$P(t)$ is simply the fraction of runs for which there is at least one replicator in the 
chain at time $t$. We stress that in the calculation of $N(t)$ we take an average over all runs,
including those that have already been extinct at time $t$, whereas $R^2(t)$ is averaged
only over the colonies that survived at time $t$. 

The idea behind the estimate of critical parameters by following the spreading of the colony 
is that  the time dependence 
differs \textit{qualitatively}  depending whether the system is 
in the supercritical or in the subcritical regime. The mere visual inspection of the
plots $N(t)$ (or $P(t)$) versus $t$ allows one to determine whether the control
parameters $D$ and $\gamma$ are below or above the critical ones.  In the case that the transition
is continuous, it is conjectured that the following scaling laws will hold 
\cite{spreading}
\begin{eqnarray}
N(t) & \sim  & t^\eta,  \label{sc1}\\
P(t) & \sim & t^{-\delta},  \label{sc2}\\
R^2(t) & \sim & t^z, \label{sc3}
\end{eqnarray}
where $\eta$, $\delta$ and $z$ are critical dynamic exponents. Particularly relevant
to our purposes is the relation between these exponents and the fractal dimension $d_f$ 
of the surviving colonies at a given asymptotically large time, namely,
$d_f = 2 \left ( \eta + \delta \right )/z$ \cite{spreading}, since the value of
this quantity, calculated at the steady-state with a different method, was used
by Fiore and de Oliveira \cite{Fiore} as the main criterion to distinguish between the continuous 
and the discontinuous transition.

In principle, power laws are not expected at a discontinuous transition
because correlations are of finite range and so   quantities such as the survival
probability $P(t)$ and the average number of particles $N (t)$ should decay exponentially 
with time \cite{DT}.  Earlier reports on power laws at discontinuous transitions in the two-dimensional 
Ziff-Gulari-Barshad (ZGB) model \cite{Evans}  and in  Conway's game of life \cite{Monetti}  were proven artifacts
of inadequately short-time simulations \cite{Albano,Mon_PRE}. On the other hand,
the one-dimensional Glau\-ber\--Ising model at zero temperature in a magnetic field \cite{Haye}
as well as the one-dimensional Domany-Kinzel cellular automaton \cite{DK} 
are   realizations of the  compact directed percolation (CDP) which exhibits a first-order
transition characterized by  power laws  with
the exponents $\eta = 0$, $\delta = 1/2$ and $z=1$ for $d=1$ \cite{Essam}. In d=2 and above, models in the CDP class 
are characterized by the mean-field  exponents  $\eta = 0$, $\delta = 1$ and $z=1$
\cite{Munoz}. Such  mean-field-like discontinuous 
transition was recently reported  in a   single-component, 
two-dimensional  lattice model of replicators \cite{Cardozo}. Thus 
there seems to be two possibilities only for the critical behavior at a  first-order irreversible 
transition between an active regime and an absorbing state: either there is no power-law behavior at all 
(as in ZGB and Conway's models) or the critical behavior is mean-field like (as in the abovementioned replicator 
model \cite{Cardozo} and in some monomer-monomer reaction systems \cite{Sans}). 
The features of the model that  determine which of these two alternatives will hold are still not well-understood.

\subsection{Conservative diffusive contact process} 

\begin{figure}
\begin{center}
\resizebox{0.75\columnwidth}{!}{%
  \includegraphics{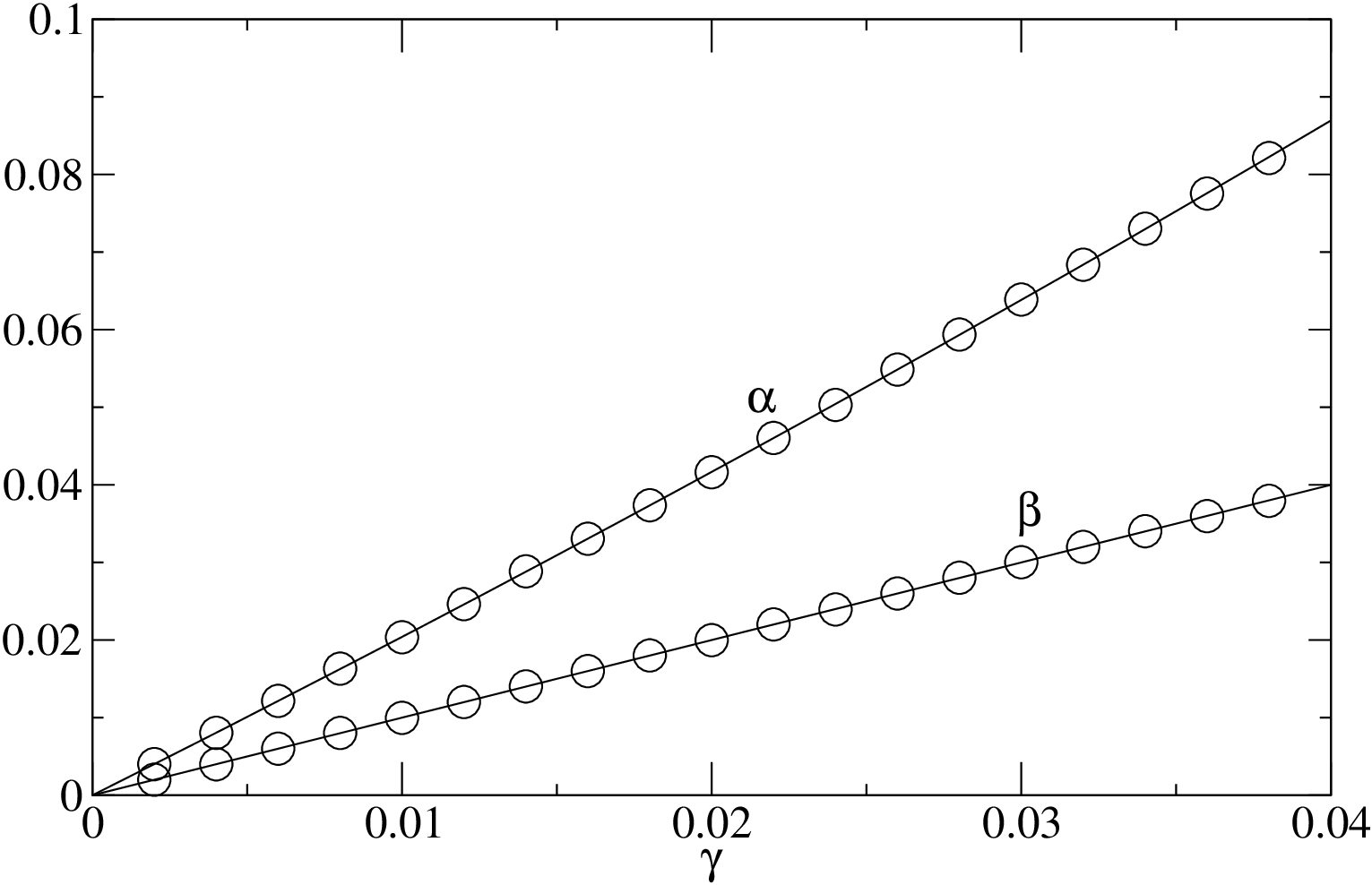}
}
\end{center}
\caption{Stationary weighted ($\beta$) and unweighted ($\alpha$) fractions of active empty sites 
as function of the decay probability $\gamma$ for the standard triplet-creation model with
cyclic boundary conditions.  The solid lines
are the fittings $\beta = \gamma$ and $\alpha = \gamma/ \left (1 - D - \gamma \right )$.
The parameters are $L = 10^4$ and $D=0.5$.}
\label{fig:1}       
\end{figure}

We now turn to the conservative diffusive contact process \cite{Fiore,CCP} in which, together
with the hopping probability $D$,
the number of replicators $N$ is kept fixed  whereas the decay
probability $\gamma$ [or, equivalently, $\lambda$, see equation (\ref{gamma})]
is derived from
the analysis of the chain configuration at the stationary state. To see how $\gamma$
can be obtained in this way we introduce the properly weighted fraction of active
empty sites,
\begin{equation}\label{beta}
  \beta = \frac{1}{N} \sum_{i} \frac{s}{2}   n_{triples}^{(i)} 
\end{equation}
where the sum is over all empty sites and $n_{triples}^{(i)}=0,1,2$ is the number of triples
adjacent to vacant site $i$.  An active empty site is a vacant site that has a nonzero
probability of being occupied. In the stationary regime the
number of replicators that decay  equals in average  the
number of replicators that are created in the active empty sites, so  one has
$ \langle  \beta \rangle = \gamma $. (The average here is over the 
distribution of occupied and vacant sites
in the steady-state regime.) We have verified the correctness of this relation 
by carrying out extensive simulations (see Fig.~ \ref{fig:1}) with the standard ensemble of variable 
particle number in a closed chain (i.e., using cyclic boundary conditions). 
Hence we can infer the value of $\gamma$ by measuring the average fraction
of active empty sites in the stationary regime, $ \langle  \beta \rangle $. 

There is, however, a difficulty to apply this scheme for the triplet-creation model. 
In the conservative diffusive contact process
there are two known  parameters -- the diffusion probability 
$D$ and the number of replicators $N$ --  but in order to calculate $\beta$ 
using equation (\ref{beta}) we need the value of the creation probability $s$, which depends
on $\gamma$, the quantity we ultimately set out to derive. The situation here is
fundamentally different from that of models for which the creation and the decay procedures 
are independent  so that the corresponding conservative contact process can easily be formulated 
\cite{Cardozo}. This hindrance can be circumvented by eliminating the factor $s$ from the definition 
of the fraction of active empty sites and considering,
instead, the quantity \cite{Fiore} 
\begin{equation}\label{alpha}
 \alpha = \frac{1}{N} \sum_{i} \frac{1}{2}   n_{triples}^{(i)} 
\end{equation}
where, as before, the sum is over all empty sites.   In Figure \ref{fig:1} we present the
dependence of the two average fractions $\beta$ and $\alpha$ on the decay probability $\gamma$
(for simplicity, henceforth we will omit the average symbols when referring to these
quantities), obtained through simulations of the standard (i.e., the number of particle
is free to vary according to the creation and decay procedures) triplet-creation model with cyclic boundary
conditions. The purpose of this figure is merely to illustrate the correctness of the
relation $\gamma = \beta$. Since from equations (\ref{beta}) and
(\ref{alpha}) we have $\beta = s \alpha$  we can easily derive the relation between $\gamma$ and $\alpha$, namely,
\begin{equation}\label{answer}
\gamma = \alpha \frac{ 1 - D }{1 + \alpha } 
\end{equation}
or, equivalently, $\lambda = 1/\alpha$.
Hence  the choice of $\beta$ or $\alpha$ is immaterial since $\gamma$ can easily be inferred from any of them.
We note in passing that the relation between the leading parameter of the conservative diffusive
contact process $\alpha$ and the
parameter of the original formulation $\lambda$ was not made explicit in \cite{Fiore}.

The basic idea of the conservative contact process is the occupation of  the active empty sites
by the transference of randomly chosen replicators to those sites, rather than by  
the creation of new replicators. In this way the
number of replicators does not change and the processes of creation and annihilation
are replaced by a single jumping process \cite{CCP}. In the conservative diffusive contact process there are two procedures
only: diffusion that occurs with probability $D$ and  jumping, which combines both creation and decay,
that occurs with probability $1-D$. Since the diffusion procedure does not change the
number of replicators, it can be implemented exactly as described in the definition of the model
\cite{Fiore}. The jumping process is implemented as follows. We choose a replicator at random and check 
whether it is at the center of a triple. If so, then  we pick at random one of the sites adjacent to the triple.
If this site is vacant we choose another replicator at random and transfer it to the vacant site. This scheme
is completely equivalent to that used by Fiore and de Oliveira \cite{Fiore}.

\begin{figure}
\begin{center}
\resizebox{0.75\columnwidth}{!}{%
  \includegraphics{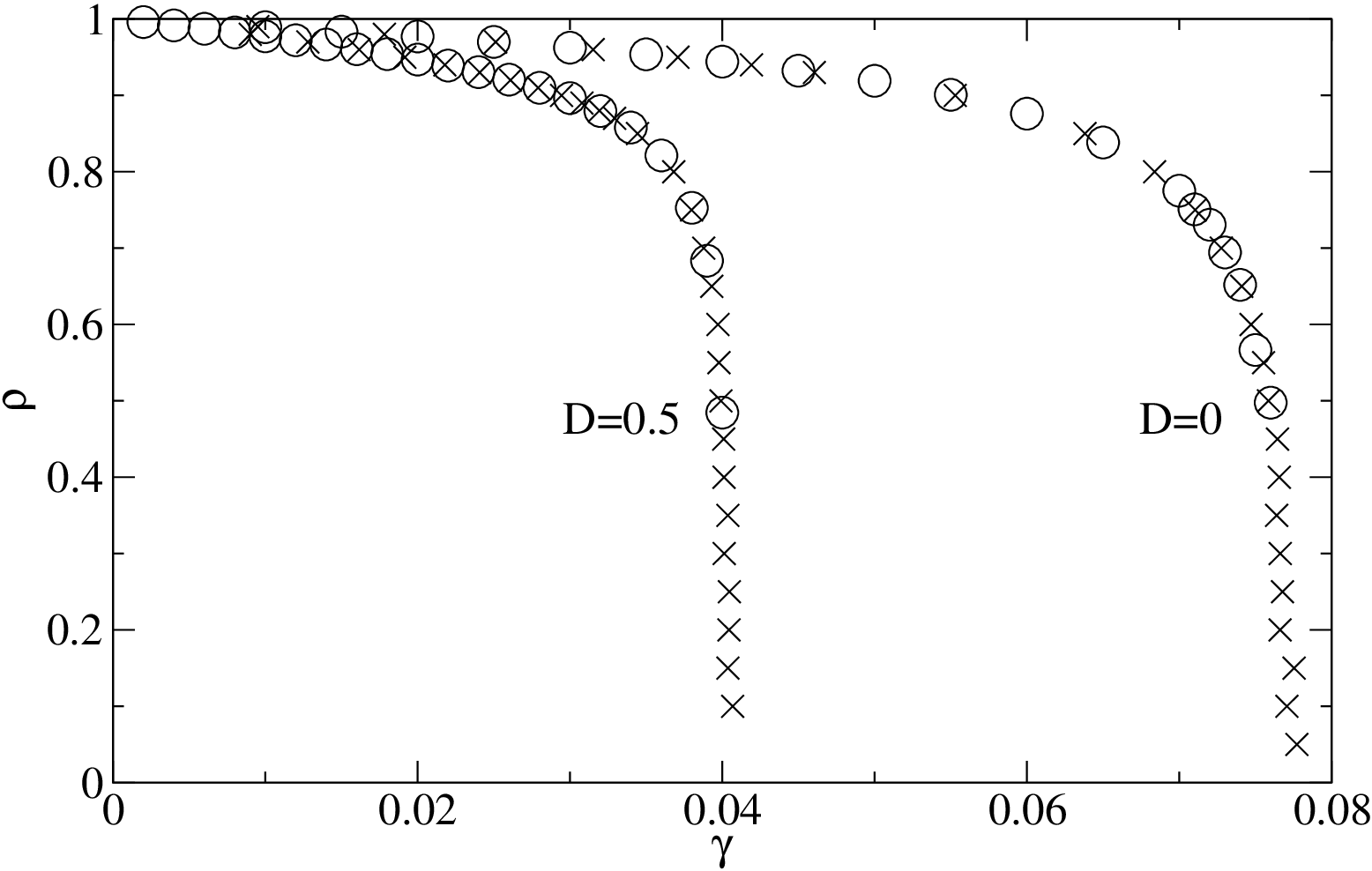}
}
\end{center}
\caption{Average density of particles as function of the decay probability of the standard 
($\bigcirc$)
and the conservative ($\times$) versions of the triplet-creation model for 
diffusion probabilities as indicated and chain size  $L = 10^4$.
}
\label{fig:2}       
\end{figure}

In Figure \ref{fig:2} we present the  comparison of the average density of particles at the
stationary state obtained with a single sample of the traditional (T) and the conservative (C) versions
of the triplet-creation model  for two values of the diffusion probability. 
At first sight the excellent agreement
between these two models with very different rules for the creation and decay processes is
truly remarkable. (We note that, leaving diffusion aside,  creation occurs with probability
$s$ in the traditional formulation,  whereas it occurs with probability one in the conservative formulation.)
But the reason for that is actually quite prosaic (see \cite{Mario} for a formal argument). For a given value of $N$, 
model C leads to a stationary state
characterized by a particular value of $\alpha$, as defined in equation (\ref{alpha}). The problem is to find
the value of $\gamma$ (or $\lambda$) in model T that produces a stationary state characterized by exactly the same value 
of $\alpha$, though these stationary states  may differ in many other aspects as, for instance, in 
the fluctuations of the number
of particles.
In view of equation (\ref{beta}) and of the fact that $\gamma = \beta$ at the stationary state of model T the
answer is  given simply by equation (\ref{answer}). Hence the agreement between models T and C, at least with regard 
to properties related to $\alpha$, follows  trivially from this argument.

In addition, Figure \ref{fig:2} provides a good  illustration of 
the difficulty to carry out simulations with the traditional ensemble very near the 
transition point since the instability of the active state for long runs in finite lattices
leads most of the samples to fall into the absorbing state. This hindrance, however, 
can be avoided by using an ensemble in which the number of replicators
is kept fixed, this being the motivation for the proposal of the conservative contact process  
\cite{CCP}.

\section{Results} \label{sec:Results} 

From the qualitative aspect, Figure \ref{fig:2} already demonstrates the main  effect
of diffusion in the model. Diffusion disrupts the triplets so that the active phase can be maintained
only by increasing the probability of creation $s$, or equivalently, by decreasing the
 probability of decay $\gamma$ as shown in the figure. In the following we will concentrate on
three values of the diffusion probability: $D=0$, for the purpose of comparison only  since
there is no dispute that the continuous transition in this
case is in the directed percolation universality class; $D=0.95$, which corresponds to
$\tilde{D} = 0.905$, roughly the value for which Hinrichsen \cite{Haye} has run  the simulations
that led to the conclusion that the transition of the triplet-creation model 
is continuous in opposition to the findings
of  Dickman and Tom\'e \cite{DT}; and $D=0.98$, which corresponds to $\tilde{D} = 0.961$
for which the conservative diffusive approach  predicts a discontinuous transition
\cite{Fiore}.

\begin{figure}

\vspace*{.5cm}

\begin{center}
\resizebox{0.75\columnwidth}{!}{%
  \includegraphics{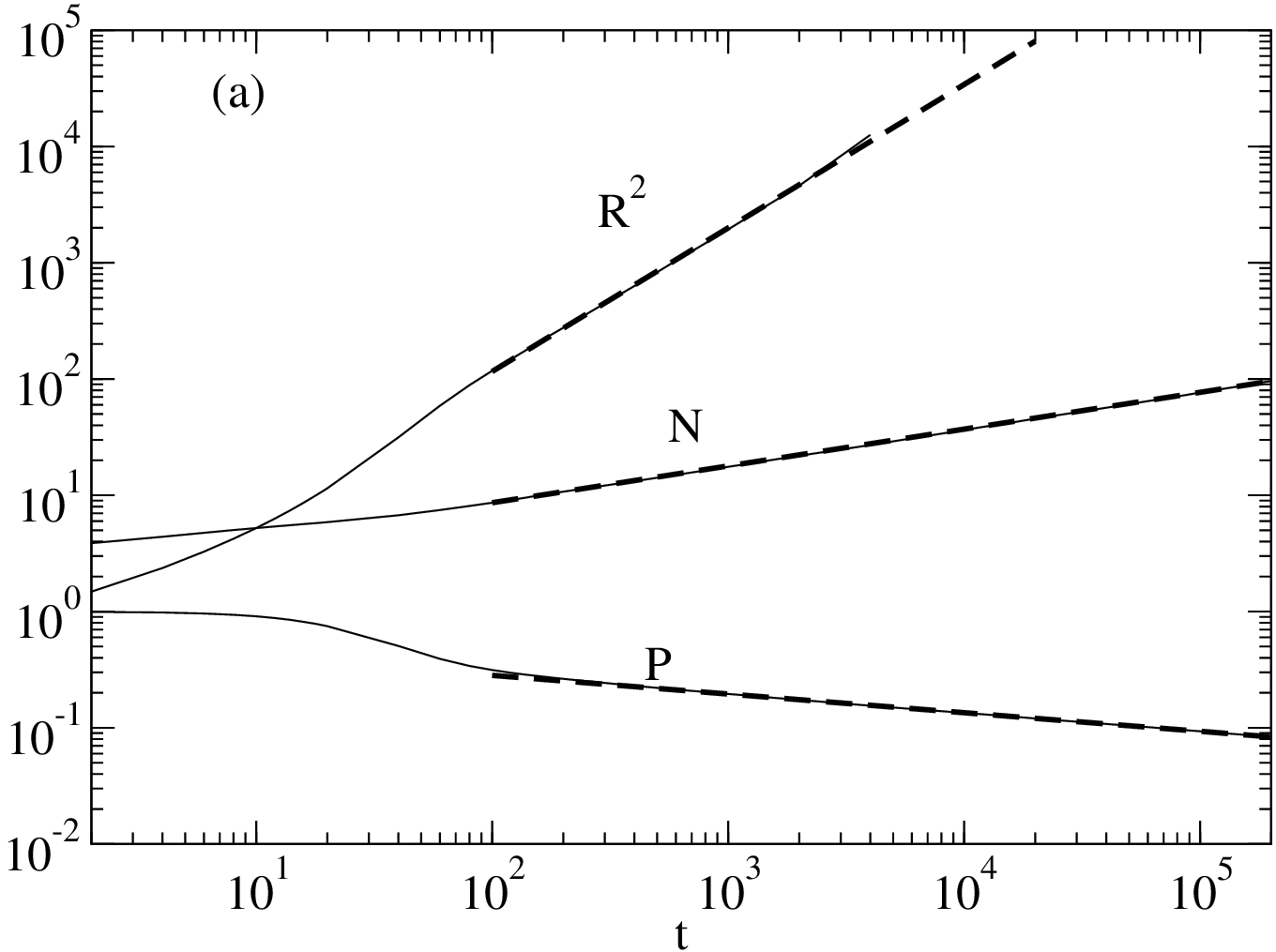} 
}  

\vspace{.8cm}

\resizebox{0.75\columnwidth}{!}{%
  \includegraphics{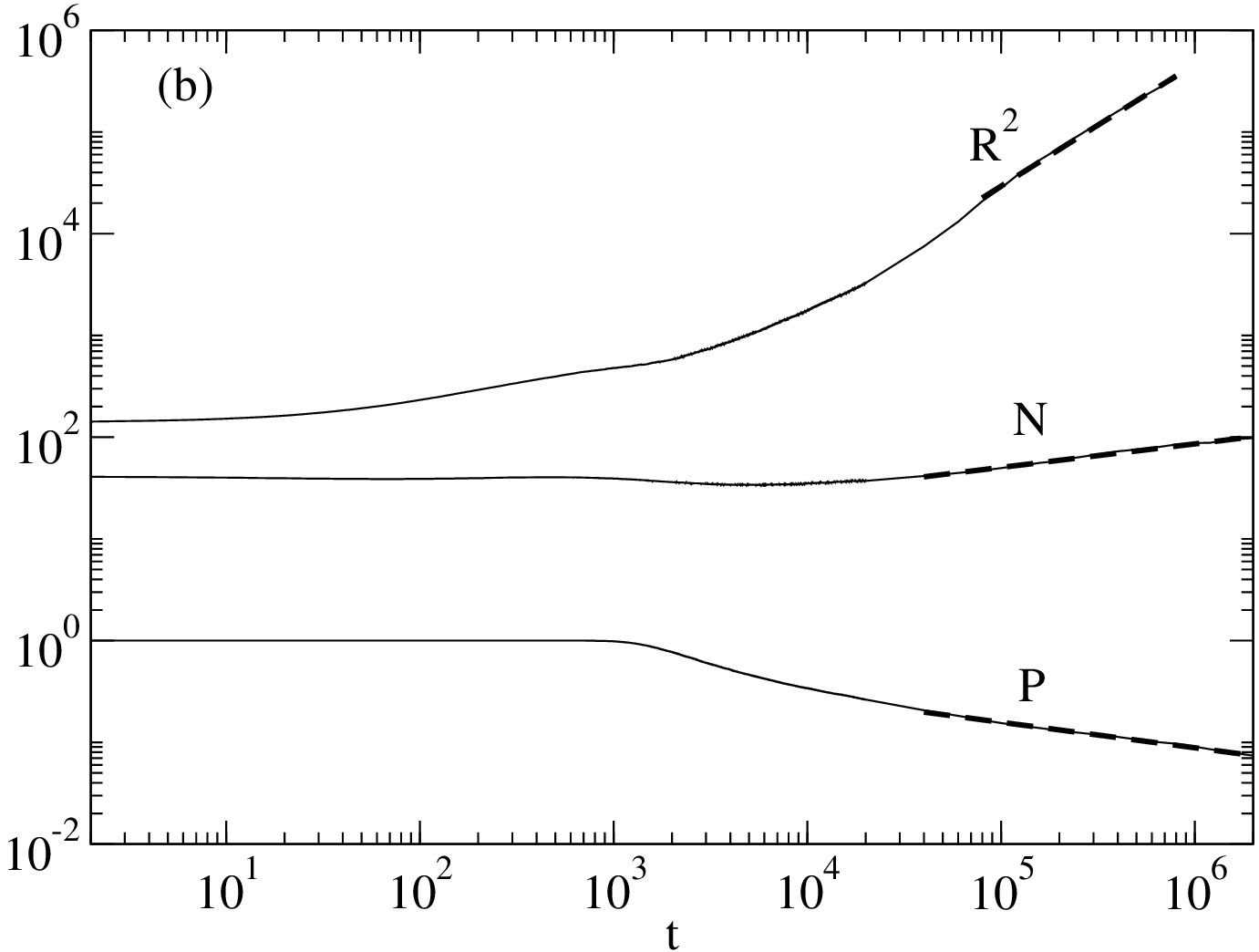}
} 

\vspace{.8cm}

\resizebox{0.75\columnwidth}{!}{%
  \includegraphics{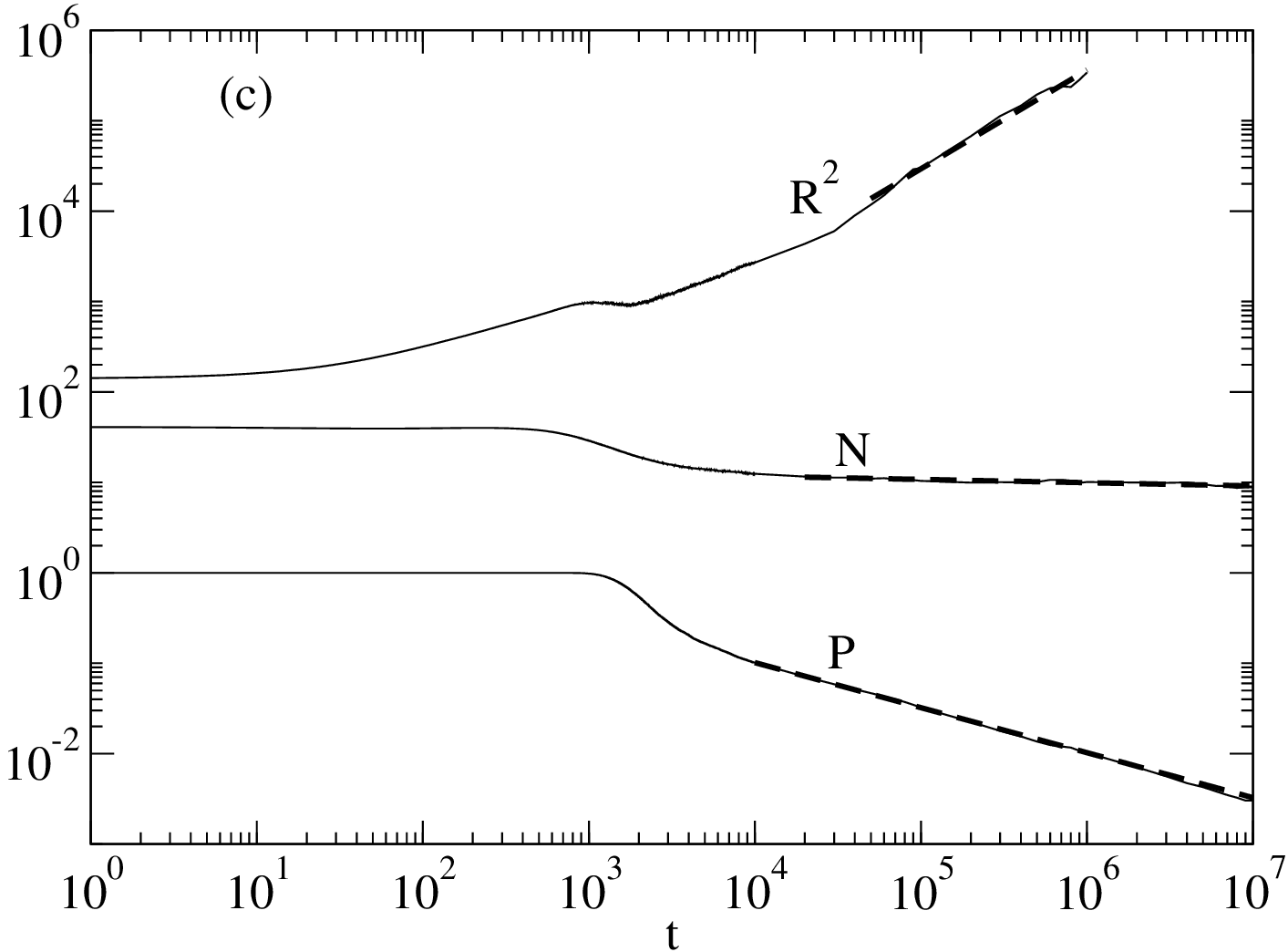}
}
\end{center}  
\caption{Results of the spreading analysis at the transition points $\gamma_c$ for
(a) $D=0$ and $\gamma_c = 0.07683(3)$; (b) $D=0.95$ and $\gamma_c = 0.00450(1)$; and (c)
$D=0.98$ and $\gamma_c = 0.001886(2)$. The numbers in parentheses
represent the uncertainty of the last digit. The dashed lines are the fittings with the scaling laws 
(\ref{sc1})-(\ref{sc3})}
\label{fig:3}       
\end{figure}

The results of the spreading analysis are presented in Figure \ref{fig:3}. As already said, for $D=0$
one recovers the dynamic exponents of the directed percolation $\eta = 0.322(5)$, $\delta = 0.164(5)$,
and $z = 1.24(2)$, which yield $d_f = 0.78(1)$.
(Henceforth the numbers in parentheses will represent the statistical uncertainty in the last digits.)
To avoid possible corrections to scale, 
the dynamic exponents were calculated using the method of the local slopes \cite{Dick}. 
The main point of Figure \ref{fig:3}a is to show the easy with which the asymptotic power-law
behavior is reached in this case: mere $10^3$ time steps are sufficient to evaluate the dynamic critical exponents
within a reasonable precision. The situation, however,  is very different for high values of the diffusion
probability. For instance, Figure \ref{fig:3}b shows the results for  $D=0.95$, the diffusion rate considered 
in the numerical analysis of Ref. \cite{Haye}. The critical spreading is characterized by the exponents
$\eta = 0.258(4)$, $\delta=0.266(6)$, and $z=1.35(2)$ which, surprisingly,
lead to the same value of the fractal dimension obtained in the fixed-position limit, $d_f = 0.78(1)$.
The initial period of leveling off in the evolution of $P(t)$ is an effect of the large initial colony
which guarantees survival in  the first $10^3$ time steps. After this initial stage, $P(t)$ enters
a regime of exponential decay which eventually crosses over to a power-law decay. The critical exponents
we find differs substantially from the DP exponents reported in Ref. \cite{Haye}. Finally, Figure \ref{fig:3}c 
shows the critical spreading curves for  $D=0.98$ which are characterized by  $\eta = -0.026(6)$, 
$\delta=0.49(2)$, and $z=1.02(1)$,
resulting in $d_f = 0.91(5)$. This result
provides strong evidence
for a discontinuous transition since these exponents are very close
to those of the CDP ($\eta =0$, $\delta=1/2$, and $z=1$). 

\begin{figure}
\begin{center}
\resizebox{0.75\columnwidth}{!}{%
  \includegraphics{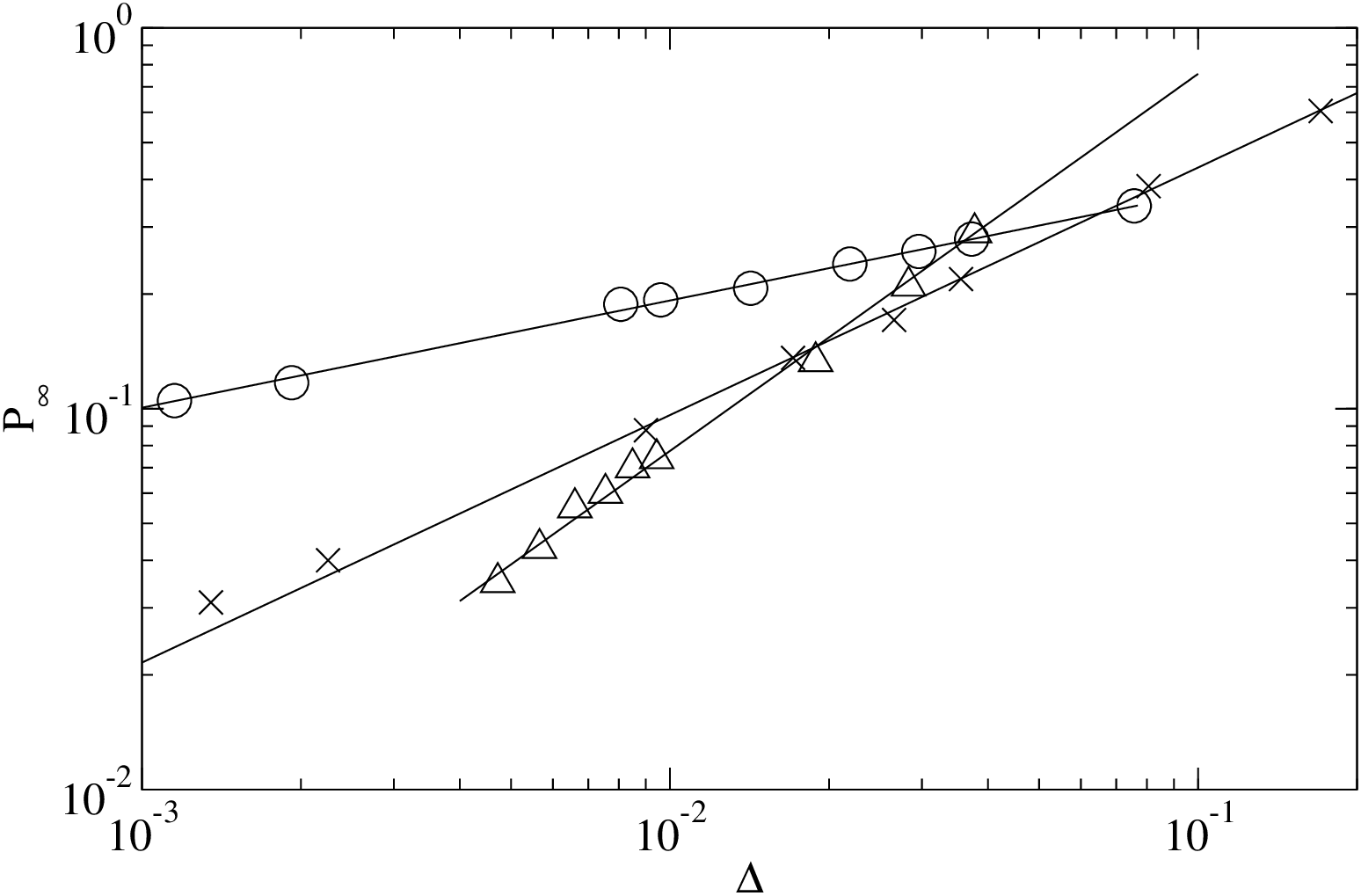}
}
\end{center}
\caption{Logarithm plot of the ultimate survival probability $P_\infty$ as function of the distance to the 
transition point $\Delta = 1 - \gamma/\gamma_c$ for  $D=0$ ($\bigcirc$), $D=0.95$ ($\times$), and $D=0.98$
($\bigtriangleup$). The straight lines are the fittings
from which the exponent $\beta'$ is calculated. }
\label{fig:4}       
\end{figure}

In addition to the
three dynamic exponents, the spreading analysis applied to the supercritical regime
($\gamma < \gamma_c$)
 permits the calculation of the exponent
$\beta'$ which controls  the approach to the critical point of the ultimate survival
probability $P_\infty = \lim_{t \to \infty} P(t)$, namely, $P_\infty \sim \Delta^{\beta'}$ 
where $\Delta = 1 - \gamma/\gamma_c$. Figure \ref{fig:4} summarizes  the results. 
We find $\beta'= 0.282(6)$ for $D=0$, $\beta'= 0.65(1)$ for $D=0.95$,
and $\beta'= 0.99(3)$ for $D=0.98$. As before, the exponents for $D=0$ and $D=0.98$ are in
excellent agreement with those of the  DP and CDP, respectively, but for $D=0.95$ the exponent
$\beta'$ settles  to a value intermediate to those extremes.

\begin{figure}

\vspace*{.5cm}

\begin{center}
\resizebox{0.75\columnwidth}{!}{%
  \includegraphics{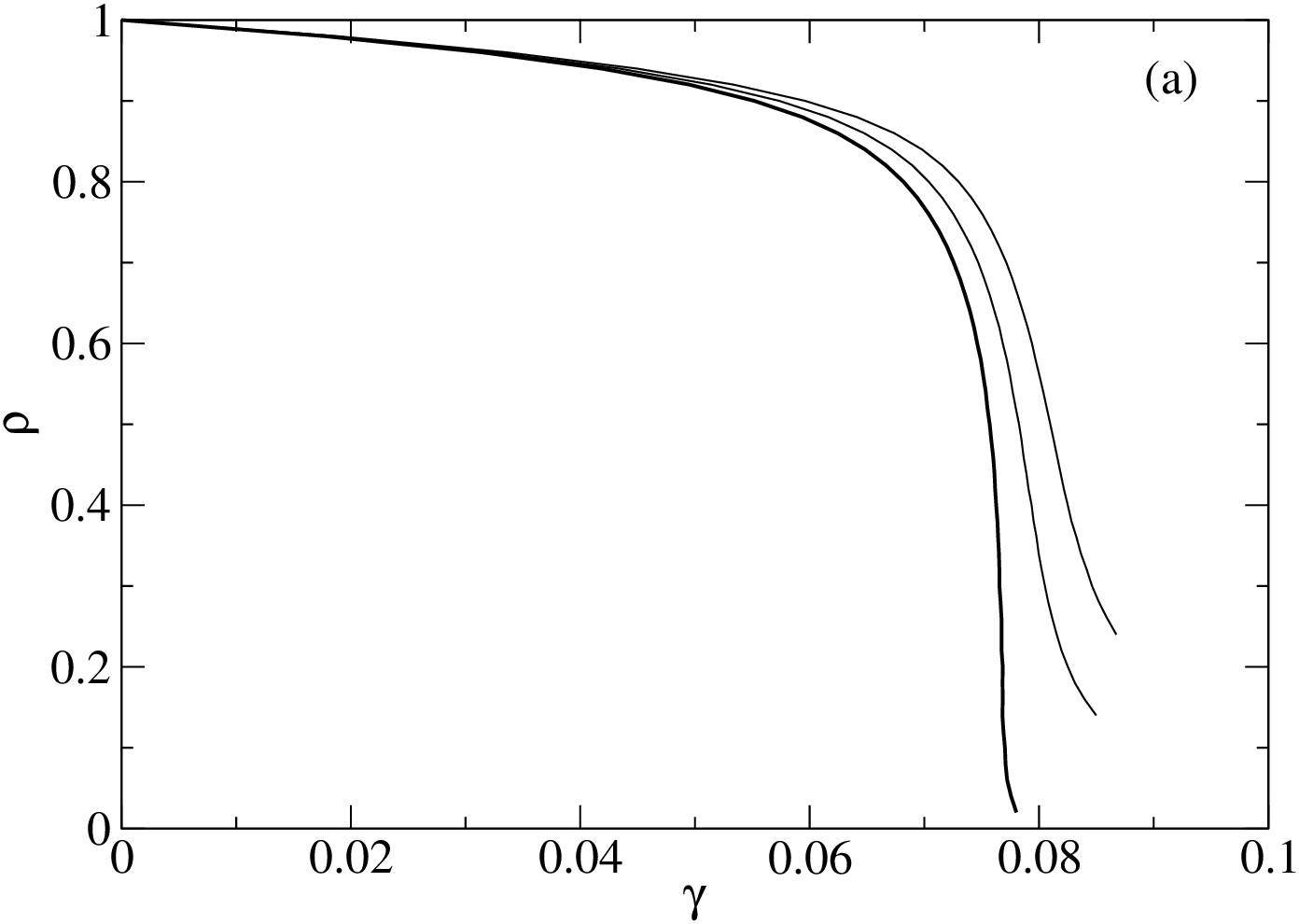} 
}  

\vspace{.8cm}

\resizebox{0.75\columnwidth}{!}{%
  \includegraphics{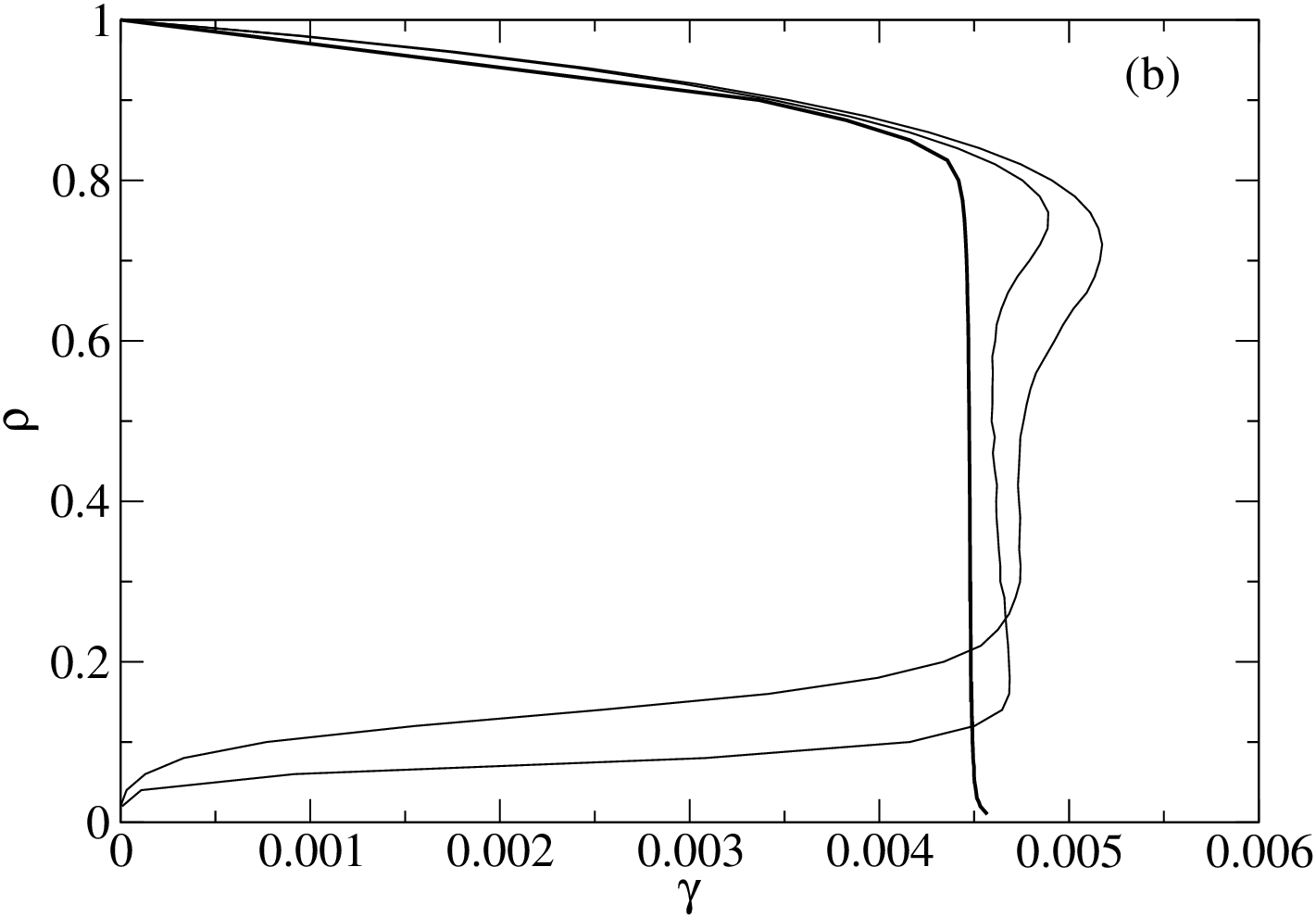}
} 

\vspace{.8cm}

\resizebox{0.75\columnwidth}{!}{%
  \includegraphics{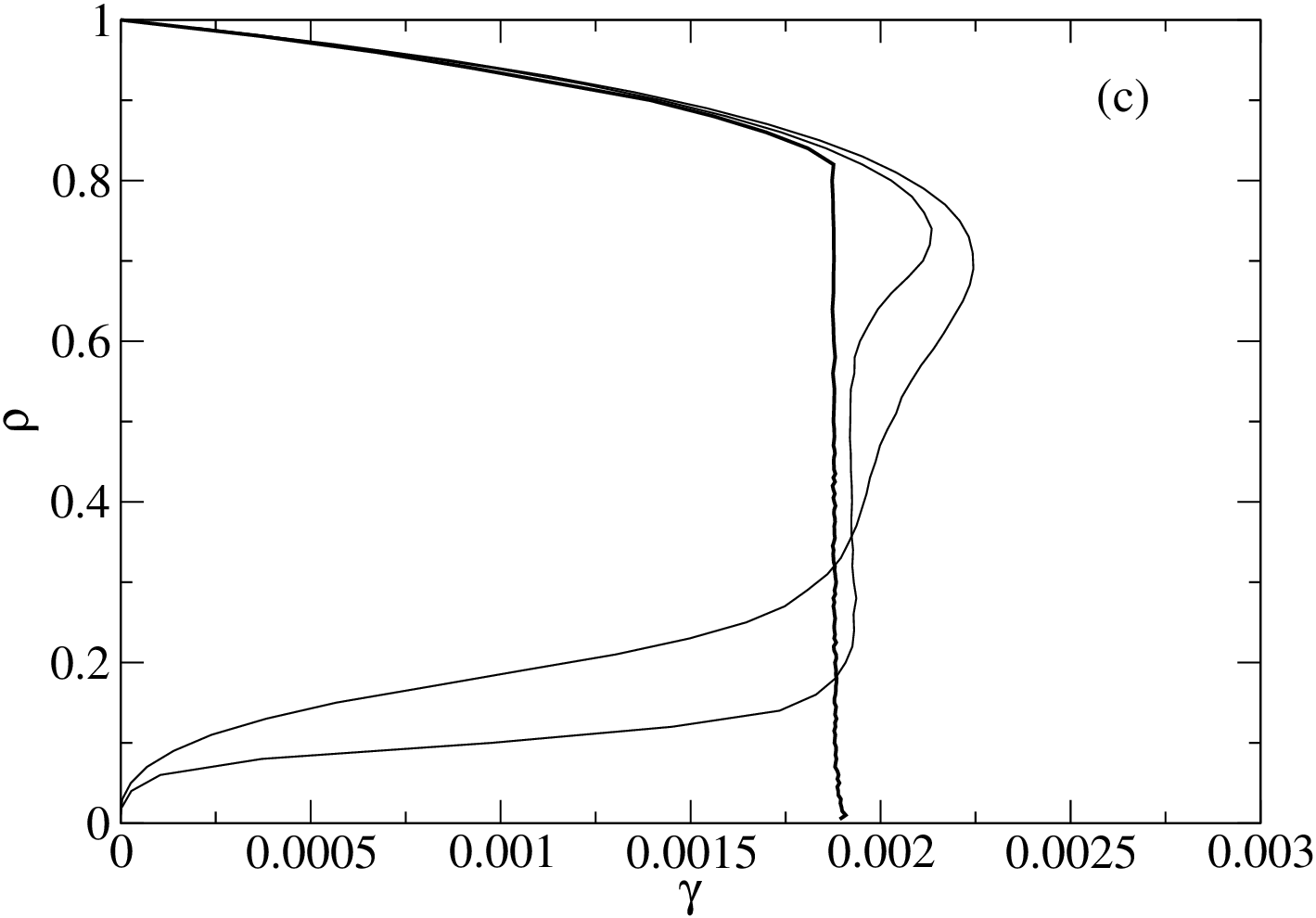}
}
\caption{Density of particles as function of the probability of decay in the
conservative diffusive contact process for 
different chain sizes (left to right at $\rho =0.6$): $L =10^4$ (thick line), $L=200$, and $L=100$. The panels are
(a) $D=0$,  (b) $D=0.95$, and (c) $D=0.98$. }
\label{fig:5}   
\end{center}      
\end{figure}

\begin{figure}

\vspace*{.5cm}

\begin{center}
\resizebox{0.75\columnwidth}{!}{%
  \includegraphics{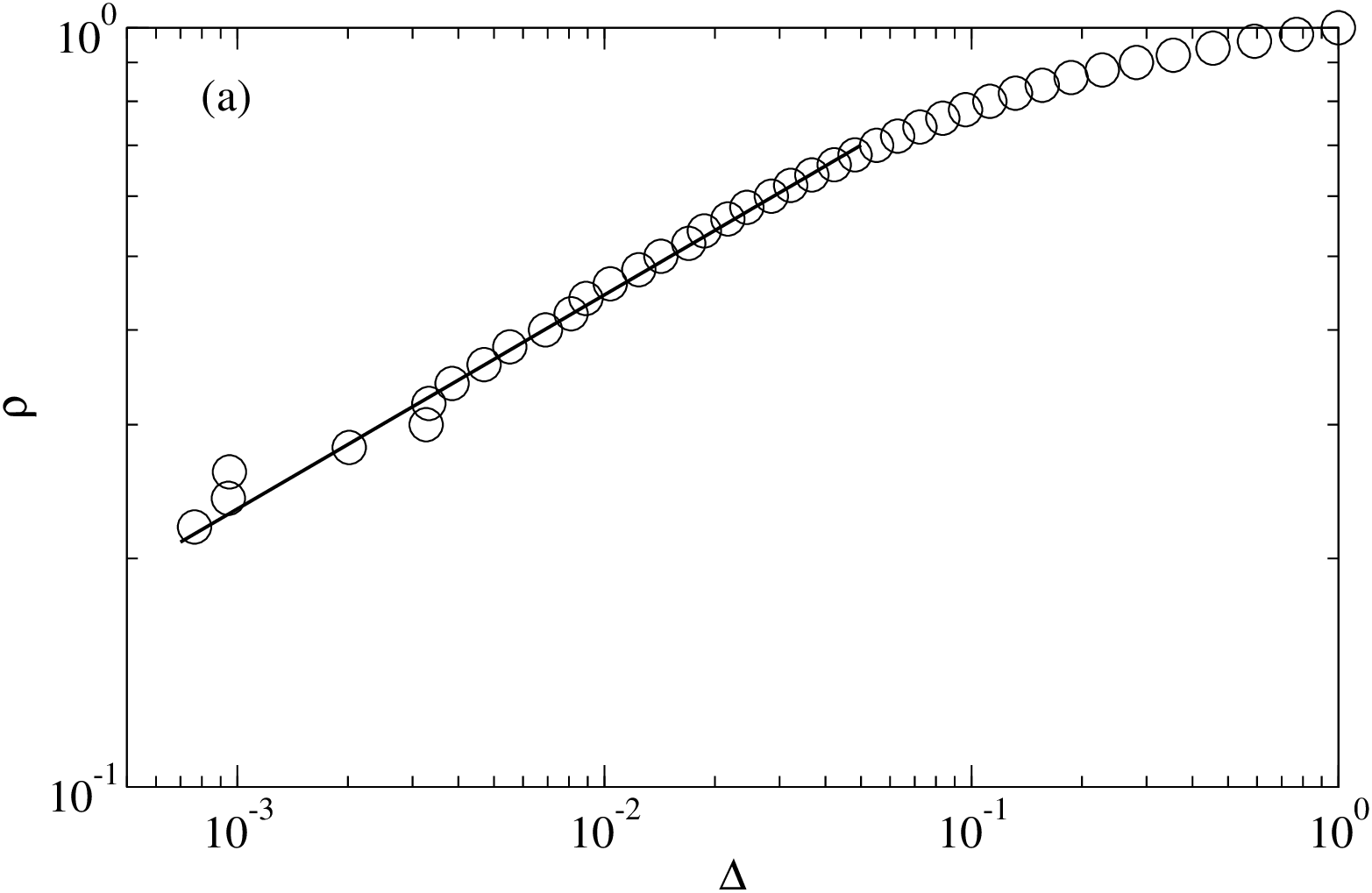} 
}  

\vspace{.8cm}

\resizebox{0.75\columnwidth}{!}{%
  \includegraphics{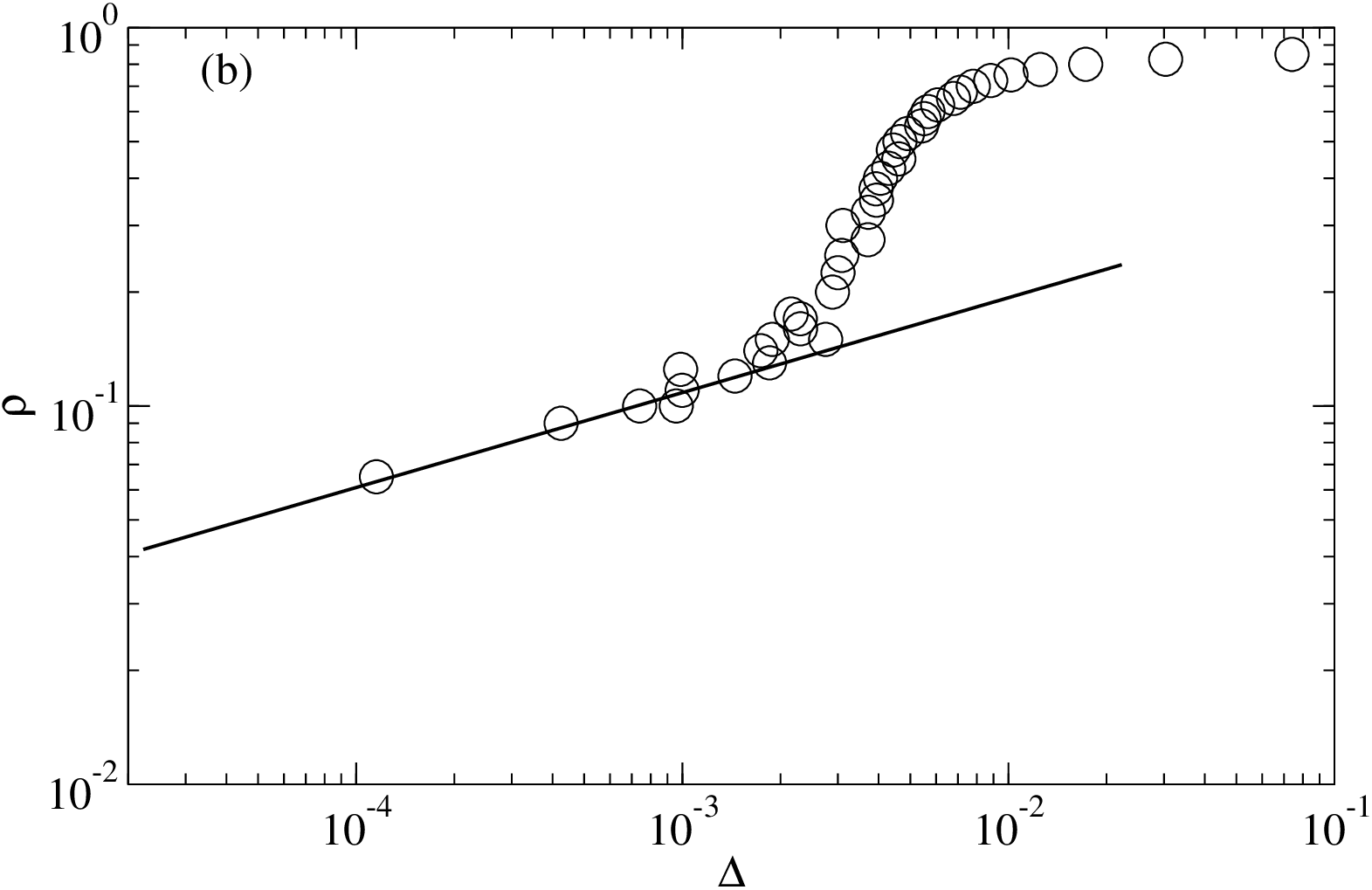}
} 

\vspace{.8cm}

\resizebox{0.75\columnwidth}{!}{%
  \includegraphics{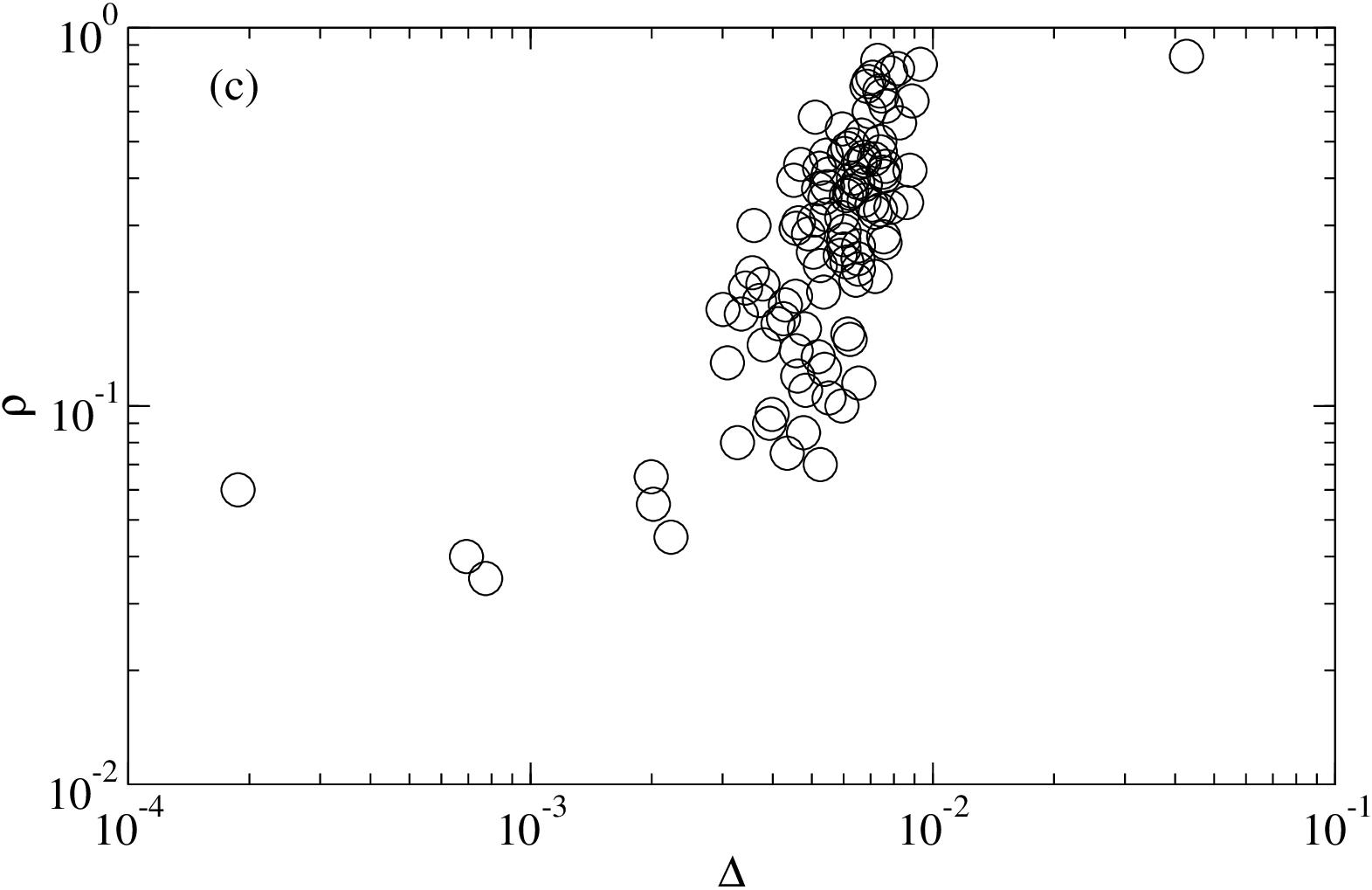}
}
\caption{Logarithm plot of $\rho$ as function of the distance to the 
transition point $\Delta = 1 - \gamma/\gamma_c$ for $L= 10^4$ and
(a) $D=0$,  (b) $D=0.95$, and (c) $D=0.98$. The straight lines are the fittings
from which the exponent $\beta$ is calculated.}
\label{fig:6}   
\end{center}      
\end{figure}

A qualitative picture of the outcome of the conservative diffusive contact process is
attained by inspection of Figure \ref{fig:5} that shows the particle density $\rho$ as
function of the decay parameter
for different chain sizes.
Note that if the non-monotonic dependence of
$\rho$ on $\gamma$ for small chains or the apparent steepness of the transition region
are used as indicators of a first-order transition, then  the transition for $D=0.95$,
which clearly exhibits these features, should be considered as discontinuous. A more
careful analysis focusing on the close vicinity of the transition point
$\gamma_c$ is presented in Figure \ref{fig:6}. Assuming that the density of particles
vanishes as $\rho \sim \Delta^\beta$  we find
 $\beta = 0.282(5)$ for $D=0$, as expected, and $\beta = 0.25(5)$  for $D=0.95$. As illustrated
in the figure, the data
for $D=0.98$ are not amenable to this kind of fitting and then we assume $\beta = 0$. 

Since the four critical exponents must obey the generalized hyperscaling relation \cite{Mendes}
\begin{equation}
h = \left ( 1 + \beta/\beta' \right ) \delta + \eta - d z/2 = 0
\end{equation}
with $d=1$, we can easily evaluate the consistency of the set of exponents  for each
value of the diffusion rate. We find
$h=0.03$ for $D=0$, $h = -0.05$ for $D=0.95$, and $h = -0.04$ for $D=0.98$ which indicate
that, despite the unusual values of the exponents, the  results for  intermediate
diffusion $D=0.95$ are as reliable as those for the undisputable limiting situations
of very low and very high diffusion probabilities.

\section{Conclusion} \label{sec:Conclusion} 

Our results corroborate the original findings of Dickman and Tom\'e \cite{DT}: the triplet-creation model
exhibits a discontinuous transition in the high diffusion regime. In stark contrast with that work, however,
we use as indicator of  the discontinuous transition not the absence of power-law decay at the
critical point, but rather the presence of a power-law behavior  characterized by the exponents of the compact directed percolation
(CDP)
$\eta =0$, $\delta=1/2$, $z=1$, and $\beta'=1$. In fact, while we recovered these exponents for high values of the
diffusion probability ($D=0.98$), we found strong crossover effects  for not so high values of this parameter ($D=0.95$), leading to
a seemingly continuous variation of the critical exponents. We note that for $D=0.9$ (data not shown) we have found the exponents of the
directed percolation (DP). Interestingly, we found that the fractal dimension $d_f$ of the colonies is not so affected
by the crossover effects. In fact, that quantity was used by Fiore and de Oliveira \cite{Fiore} as the indicator to locate
the tricritical point.  The calculation of $d_f$ within the conservative diffusive contact process framework is based on
the scaling relation between the (fixed) number of particles $N$ and the average distance $R$ between the two particles located at the
extremities of the chain, $N \sim R^{d_f}$. In a  sense the situation here is reminiscent of models with infinitely
many absorbing states, for which only a combination of the dynamic exponents (the sum $\eta + \delta$ in that case) is universal 
\cite{Mendes,Lipowski}. Nonetheless, the sole coincidence of the values of $d_f$ for $D=0$ and $D=0.95$, whereas all other exponents
differ so markedly (e.g., $\beta'$), should be viewed as a  weak evidence, rather
than a conclusive indication,  for a continuous transition at $D=0.95$. 

The triplet-creation model provides a simple, but surprisingly difficult,
test case to study the crossover from DP to CDP.
The fact that the critical exponents seem to 
change continuously from those
of DP for $D < 0.9$ to those of CDP for $D > 0.98$ is puzzling, since what
is normally observed (and  expected) is a transient behavior in which the different universality
classes dominate 
within different time regimes \cite{Sans,crossover,Kevin}.
However, since the exponents $\beta'$, $\beta$ and the set of dynamic exponents $\eta, \delta, z$ are calculated through
independent techniques it is unlikely that, by increasing the chain size or the total evolution time, one would be
able to recover the familiar exponents of the DP or CDP universality classes. Clearly, further
research is needed to clarify the crossover from these two classes in the triplet-creation model.
Despite the interest on this
model \cite{DT,Haye,Fiore}, the systematic evaluation of  the entire set of critical  exponents was still lacking. The 
aim of this contribution was to fill this gap and, in doing so, we have unveiled a  rich crossover phenomenon whose
elucidation  will pose a hard challenge to the current techniques for characterization of nonequilibrium critical
behavior.

\bigskip

{\small The work of J.F.F. was supported in part by CNPq and FAPESP, Project No. 04/06156-3.
G.O.C. was supported by FAPESP.}


\begin{thebibliography}{99}

\bibitem{Maynard} J. Maynard Smith, E. Szathm\'ary, \textit{The Major
Transitions in Evolution} (Freeman, Oxford, 1995); \textit{The Origins of Life}
(Oxford University Press, Oxford, 1999)

\bibitem{Ferreira} C.P. Ferreira, J.F. Fontanari, Phys. Rev. E \textbf{65},  021902 (2002);
A. Rosas, C. P. Ferreira, J.F. Fontanari, Phys. Rev. Lett.  \textbf{89}, 188101 (2002);
A. Rosas, J.F. Fontanari, Orig. Life Evol. Biosph. \textbf{33},  357 (2003)

\bibitem{Michod}R. E. Michod, Amer. Zool. \textbf{23}, 5 (1983)

\bibitem{Marro} J. Marro, R. Dickman, \textit{Nonequilibrium Phase Transitions in 
Lattice Models} (Cambridge University Press, Cambridge, 1999)

\bibitem{review} H. Hinrichsen, Adv. Phys. \textbf{49},  815 (2000)

\bibitem{Odor} G. \'Odor, Rev. Mod. Phys. \textbf{76},  663 (2004)

\bibitem{spreading} P. Grassberger, A. de La Torre, 
Ann. Phys. (NY) \textbf{122},  373 (1979); P. Grassberger, J. Phys. A \textbf{22}  3673 (1989).

\bibitem{DT} R. Dickman, T. Tom\'e, Phys. Rev. A \textbf{44}, 4833 (1991)

\bibitem{Hyper} M. Eigen, P. Schuster, \textit{The Hypercycle: 
A Principle of Natural Self-Organization} (Springer-Verlag, New York, 1979)

\bibitem{Boerlijst} M.C. Boerlijst, P. Hogeweg, Physica D \textbf{48}, 17 (1991) 

\bibitem{DP} H.K. Janssen, Z. Phys. B \textbf{42}, 151 (1981)

\bibitem{Grassberger} P. Grassberger, Z. Phys. B \textbf{47}, 365 (1982); J. Stat. Phys. \textbf{79}, 
13 (1995) 

\bibitem{Haye} H. Hinrichsen, cond-mat/0006212

\bibitem{Fiore}  C.E. Fiore, M.J. de Oliveira, Phys. Rev. E \textbf{70},  046131 (2004)

\bibitem{CCP} T. Tom\'e, M.J. de Oliveira, Phys. Rev. Lett. \textbf{86},  5643 (2001);
M.M.S. Sabag, M.J. de Oliveira, Phys. Rev. E \textbf{66},  036115 (2002)

\bibitem{Mendes} J.F.F. Mendes, R. Dickman, M. Henkel, M.C. Marques, J. Phys. A \textbf{27}, 3019  (1994)

\bibitem{Dick} R. Dickman, Phys. Rev. A \textbf{42},  6985 (1990)

\bibitem{Evans} J.W. Evans,  M.S. Miesch, Phys. Rev. Lett.  \textbf{\bf 66},  833 (1991)

\bibitem{Monetti} R.A. Monetti, E.V. Albano, J. Theor. Biol. \textbf{187}, 183 (1997)

\bibitem{Albano} R.A. Monetti, E.V. Albano, J. Phys. A \textbf{34},  1103 (2001);
E.V. Albano, R.A. Monetti, Surf. Rev. Lett. \textbf{10},  867 (2003)

\bibitem{Mon_PRE} R.A. Monetti, Phys. Rev. E \textbf{65},  016103 (2001)

\bibitem{DK} E. Domany, W. Kinzel, Phys. Rev. Lett. \textbf{53}, 447 (1984) 

\bibitem{Essam} J.W. Essam, J. Phys. A \textbf{22}, 4927  (1989) 

\bibitem{Munoz} M.A. Mu\~noz, R. Dickman, A. Vespignani, S. Zapperi,
Phys. Rev. E \textbf{59},  6175 (1999)

\bibitem{Cardozo} G.O. Cardozo, J.F. Fontanari, Physica A \textbf{359}, 478 (2006)

\bibitem{Sans} M.A. Sanservino, A. L\'opez, E.V. Albano, R.A. Monetti, Eur. Phys. J. B
\textbf{40}, 305 (2004)

\bibitem{Mario} M.J. de Oliveira, Phys. Rev. E \textbf{67},  027104 (2003)

\bibitem{Lipowski} A. Lipowski, M. Lopata, Phys. Rev. E \textbf{60},  1516 (1999)

\bibitem{crossover} J.F.F. Mendes, R. Dickman, H. Herrmann, Phys. Rev. E \textbf{54}, R3071 (1996)

\bibitem{Kevin} K.E. Bassler, D.A. Browne, Phys. Rev. E \textbf{55}, 5225 (1997)



















\end{thebibliography}
\end{document}